\documentclass{iaus}
\usepackage{graphicx}

\title[Stellar Activity Observations in Argentina]
{12 Years of Stellar Activity Observations in Argentina}

\author[Pablo J.D. Mauas et al.]
{Pablo J.D. Mauas$^1$, A. Buccino$^1$, R. D\'iaz$^2$, M. Vieytes$^1$,
R. Petrucci$^1$, E. Jofre$^3$, X. Abrevaya$^1$, M.L. Luoni$^1$, P. Valenzuela$^1$} 

\affiliation{$^1$Instituto de Astronom\'ia y F\'isica del Espacio, \\ C.C. 67 - Suc. 28, 1428, Buenos Aires, Argentina
\\ email: {\tt pablo@iafe.uba.ar} \\[\affilskip]
$^2$ Institut d'Astrophysique de Paris, CNRS/UPMC, Paris, France.
Observatoire de Haute-Provence, CNRS/OAMP, Saint-Michel
l'Observatoire, France.\\[\affilskip]
$^3$ Observatorio Astron\'omico de C\'ordoba, Argentina

}

\pubyear{2008}
\volume{xxx}  
\pagerange{119--126}
\jname{Title of your IAU Symposium}
\editors{A.C. Editor, B.D. Editor \& C.E. Editor, eds.}
\begin{document}

\maketitle

\begin{abstract}
We present an observational program we started in 1999, to
systematically obtain mid-resolution spectra of 
late-type stars, to study in particular chromospheric activity. 
In particular, we found cyclic
activity in four dM stars, including Prox-Cen. 
We directly derived the conversion factor that translates the known
$S$ index to flux in the Ca~II cores, and extend its calibration to a
wider spectral range. 
We investigated the relation between the activity measurements in the 
calcium and hydrogen lines, and found that the usual correlation observed is 
the product of the dependence of each flux on stellar color, and it
is not always preserved when simultaneous observations of a particular
star are considered. 
We also used our observations to model the chromospheres of stars of
different spectral types and activity levels, and found that the
integrated chromospheric 
radiative losses, normalized to the surface luminosity, show a unique
trend  for G and K dwarfs when plotted against the $S$ index.

\end{abstract}

\firstsection 
\section{Introduction}

In 1999 we started a program to systematically obtain spectra of
late-type stars, to study in particular chromospheric activity. The
stars were chosen to cover the spectral range from F to M, with
different activity levels. Since we were particularly interested in
the transition to completely convective stars, we included a larger
number of M stars in our sample.   

Our observations were made at the 2.15~m telescope of the Complejo
Astron\'omico El Leoncito (CASLEO), which is located at 2552~m above
sea level, in San Juan, Argentina. We obtained high-resolution 
echelle spectra with a REOSC spectrograph. The maximum
wavelength range of our observations is from 3860 to 6690~\AA, 
and the spectral resolution ranges from 0.141 to 0.249~\AA\
per pixel ($R=\lambda / \delta \lambda \approx 26400$).

At present, we have about 5500 spectra of 150 stars, ranging from F to
M with different activity levels. Altough most of the stars are single
dwarfs, we have several binaries and a few subgiants. Currently, we
have four observing runs per year. Details on the reduction and
calibration procedures can be found in \cite[Cincunegui \& Mauas (2004)]{CM04}.



\section{Flux-Flux calibrations}

Since, unlike most surveys of this kind, the observations in the
different spectral features are made simultaneously, our data provides
an excellent opportunity to study the correlation between different
spectral features and activity indexes. 
To date, the most common indicator of chromospheric activity is 
the well-known $S$ index, essentially the ratio of the flux in the core of the Ca 
II H and K lines to the continuum nearby (\cite{1978PASP...90..267V}). 
This index has been defined at the Mount Wilson Observatory, were an extensive 
database of stellar activity has been built over the last four decades. 
However, unlike our program, these observations are mainly
concentrated on stars ranging from F to K, due to the long exposure times 
needed to observe the Ca II lines in the red in faint M stars. For this 
reason, the $S$ index is poorly characterized for these stars. We
first obtained the $S$ index for our spectra, integrating with the
corresponding profile (for details, see \cite{CDM07}). 

The Mount Wilson $S$ index can be converted to the average surface flux in the 
Ca II lines through the relation: 
\begin{equation}
F_{\mathrm{HK}} = F_{\mathrm{bol}} 1.34\, 10^{-4}\, S\, C_{\mathrm{cf}} 
\, ,
\end{equation} 
where $C_{\mathrm{cf}} \left( B\!-\!V\right)$ is a conversion factor that
depends on color. Two different expressions are widely used for this 
factor, the first one given by \cite{1982A&A...107...31M} and corrected by 
\cite{1984ApJ...279..763N} and the other one given by 
\cite{1984A&A...130..353R}. 
The deductions used in both works to derive $C_{\mathrm{cf}}$ involve complex 
calibration procedures. 

\begin{figure}\centering
\includegraphics[width=7cm]{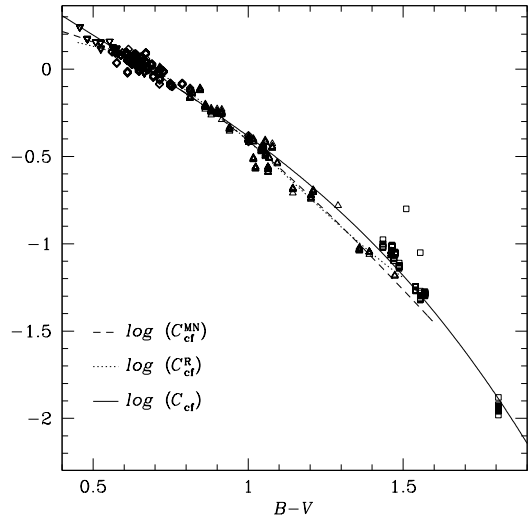}
\caption{Conversion factor between Mount Wilson $S$ index and $F_{\mathrm{HK}}$. 
The dashed line corresponds to \cite{1984ApJ...279..763N}, the dotted line to 
\cite{1984A&A...130..353R}, and the full line to our derived factor
(from \cite{CDM07}).}
\label{CC}
\end{figure}

Since we have simultaneous measurements of the $S$ index and the core 
fluxes, we can calculate directly the correction factor as a function of the 
index and the flux, which is shown in Fig. \ref{CC} together with the
other two expressions (for details and a whole discussion, see \cite{CDM07}).

It is usually accepted that there is a tight relation between the 
chromospheric fluxes emitted in H-$\alpha$ and in the H and K Ca II
lines, and that these two features can be used to study chromospheric
activity. However, most works 
where this relation has been observed  found it by using 
averaged fluxes for both the calcium and the hydrogen lines, which were not 
obtained simultaneously, and were even collected from different
sources. 

\begin{figure}\centering
\includegraphics[width=10cm]{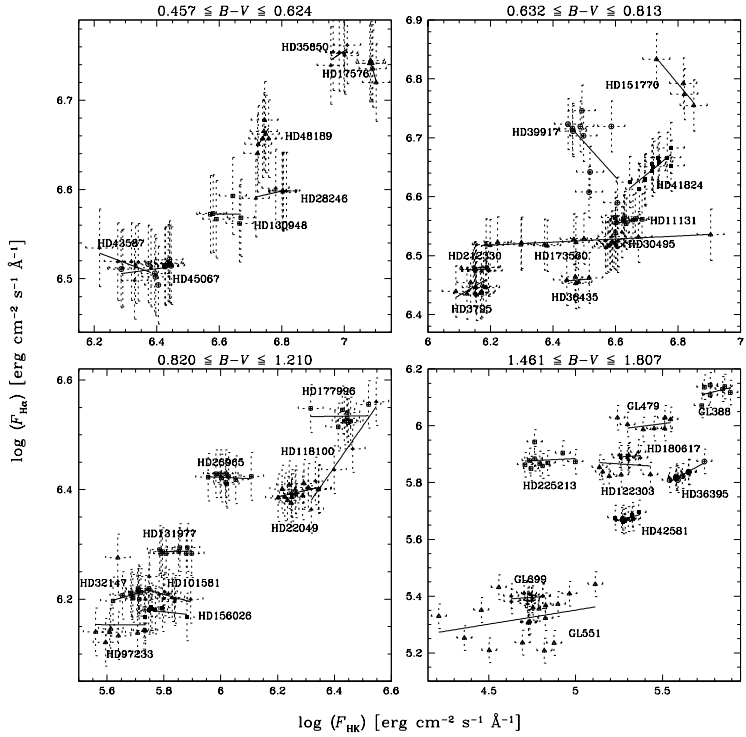}
\caption{H-$\alpha$ vs. Ca II 
surface fluxes, for stars of different spectral types, divided into different 
color bins, as indicated (from \cite{CDM07}).} \label{bins}
\end{figure}

However, the situation is different when the individual 
stars are studied separately. In Fig.~\ref{bins} we plot the 
individual simultaneous measurements of each flux for several stars of 
different spectral types and different levels of activity, divided into color 
bins as stated. We also show 
the linear fits for each star. It can be seen that the behavior is different in 
each case: in some stars both fluxes are well correlated,  although
the slopes of the fits are not the same. In other stars the H-$\alpha$
flux seems to be almost independent of the level of  
activity measured in the Ca II lines, and there are even stars 
where the fluxes are anti-correlated.

In \cite{DCM07} we studied the sodium D lines (D1: 5895.92 \AA; D2: 5889.95
\AA) in our stellar sample.  
We found a good correlation between the equivalent width of the D lines
and the color index $(B-V)$ for all the range of observations. 
Since equivalent width is a
characteristic of line profiles that do not require high resolution
spectra to be measured, this fact could become a useful tool for
subsequent studies. Finally, we constructed a spectral index ($R^\prime_D$) as the
ratio between the flux in the D lines and the bolometric flux. Once
corrected for the photospheric contribution, this index can be used as
a chromospheric activity indicator in stars with a high level of
activity. Additionally, we found that combining some of our results, we
obtained a method to calibrate in flux stars of unknown color.


\section{Cycles in M-stars}

\begin{figure}\centering
\includegraphics[width=7cm]{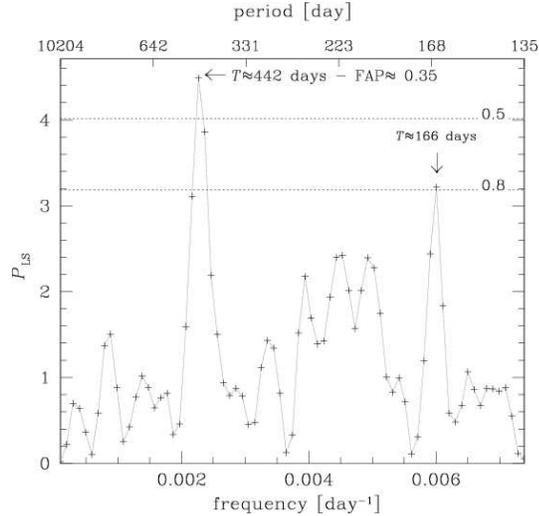}
\caption{Lomb-Scargle periodogram of our observations. The 
False Alarm Probability levels of 50 and 80\% are shown.
(from \cite{prox07}).}
\label{prox-per}
\end{figure}

One of the main goals of our program was to extend the studies of
stellar cycles to M-stars, at and beyond the limit for full
convectivity. The first star we studied was Proxima Centaury, a dMe
5.5 star with strong and frequent flaring activity \cite[Cincunegui, D\'iaz \& Mauas (2007b)]{prox07}.
For this star we excluded the spectra taken during flares, computed
the nightly average of the H$\alpha$ flux, 
and calculated the Lomb-Scargle periodogram, which is shown in
Fig. \ref{prox-per}. In the periodogram we found strong evidence of a
cyclic activity with a period of $\sim$442 days. Similar values for the 
period were found using three different techniques in the time domain
(see \cite{prox07} for details). We were also able to determine that
the activity variations outside of flares amount to 130\% in $S$,
three times larger than for the Sun.

Since this star should be fully convective, it cannot support an
$\alpha\Omega$ dynamo, and a different mechanism should be found 
to explain this result.
Recently,  \cite{CK06} showed that these objects can 
support large-scale magnetic fields by a pure $\alpha^2$ dynamo process.
Moreover, these fields can produce the high levels of activity observed 
in M stars (see, for example, \cite{ADL96}.
This $\alpha^2$ dynamo does not predict a cyclic activity. 
However, our observations suggest that this cool star has a clear period. 

We also studied the spectroscopic binary system Gl 375 (\cite{gl375}).
We first obtained precise measurements of the orbital period (P = 1.87844
days) and separation (a = 5.665 $\mathrm{R_\odot}$), minimum masses
and other orbital
parameters. We separated the composite spectra into those
corresponding to each component, which allows us to confirm that both
components are of spectral type dMe 3.5.

To study the variability of Gl~375, besides using the spectra obtained
at CASLEO, we also employed photometric observations provided by the
All Sky Automated Survey (ASAS \cite{pojmanski2002}). We calculated
the Lomb-Scargle periodogram for these data, and obtained a a distinct
peak corresponding to a period of $\mathrm{P_{phot}} = 1.876667$ days,
a period resembling very closely the measured orbital period.
We believe this harmonic variability is produced by spots and
active regions in the stellar surface carried along with
rotation. This would imply that the rotational and orbital periods are
synchronized, as is expected for such a close binary.

\begin{figure}\centering
\includegraphics[width=12.5cm]{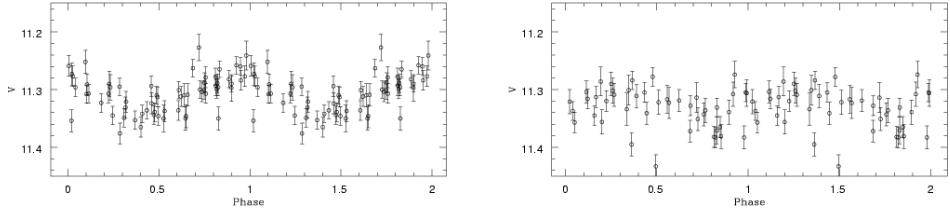}
\caption{ASAS photometry phased to the orbital period for two different epochs. Left: 2002.5-2003.5 Right: 2006.
(from \cite{gl375}).}
\label{ASAS}
\end{figure}

To verify if this is indeed the case, we phased the data to 
the obtained period, for two different seasons (Fig.~\ref{ASAS}).
The sinusoidal shape of the variation is evident, although the 
amplitude of the modulation is different in both cases, probably an
indication of 
different area covered by starspots or active regions, and 
therefore different activity levels, in each epoch. Therefore, the
amplitude of this modulation can be used as an activity proxy, and
indicates that the system exhibits a
roughly periodic behavior of 2.2 years (or ∼800 days). The
same period was found in the mean magnitude of the system and
in the flux of the Ca II K line, although the Ca flux variations
occur 140 days ahead of the photometric ones, a behavior that
has been previously observed in other stars (\cite{Gray96}). 
The agreement between the behavior of the three observables
is remarkable because of the different nature of the observations
and the different instruments and sites where they
were obtained.

\begin{figure}\centering
\includegraphics[width=7cm]{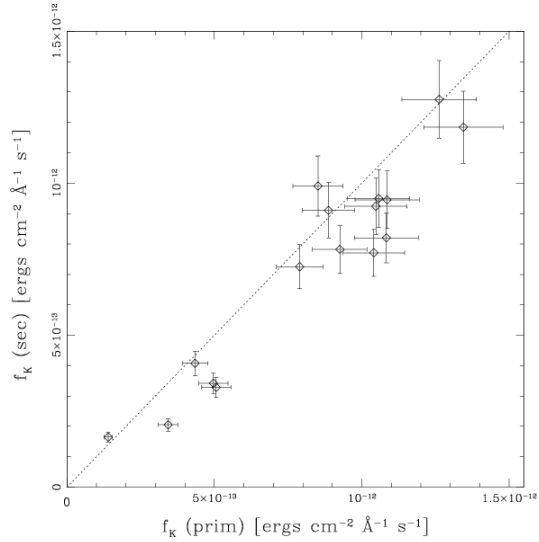}
\caption{Comparison of the fluxes in the Ca II K line for both
components. The error bars correspond to a 10\% error in the line
fluxes, and the dotted line is the identity relation. 
(from \cite{gl375}).}
\label{gl375twocomp}
\end{figure}

Another interesting result of this work is that the activity
of Gl 375 A and Gl 375 B, as measured in the flux of
the Ca II K lines, are in phase, as can be seen in
Fig. \ref{gl375twocomp}. There is an excellent correlation between the 
levels of chromospheric emission of both components,
implying a magnetic connection between them.
Due to its vicinity and relative brightness, this system
presents an interesting opportunity to further study this type
of interaction.

We also studied the long-term activity of two other M dwarf stars:
Gl 229 A and Gl 752 A, using again the Ca II - K 
line-core fluxes measured on our spectra and the ASAS photometric data.
Using the Lomb-Scargle periodogram, we obtained a possible activity cycle of
$\sim$4 and $\sim$7 yrs for Gl 229 A and Gl 752 A, respectively
(\cite{Bucetal11}). This work was complemented by other studies, were
we investigated the presence of activity cycles using IUE data
({\cite{BM08} and \cite{BM09}).

\section{Chromospheric modeling}

We also used our spectra as observational basis to model the
chromosphere of different type of stars ans activity levels, following
the path initiated with the model of Ad-Leonis in its quiescent state
(\cite{MF94}), which was completed with
a study of other dM 3.5 stars of different activity levels
(\cite{MFPP97}). We first computed models of the Sun as a star and 9
solar analogues (\cite[Vieytes, Mauas \& Cincunegui 2005]{vmc05}) which were followed by models for 6 K
stars, ``analogs'' of epsilon-Eridani (\cite[Vieytes, D\'iaz \& Mauas 2009]{vdm09}). For most stars, we built two
models, to match the observations in its minimum and maximum levels of
activity, and found that the differences in the atmospheric structure
for a star in its maximum and minimum activity levels are similar to
the changes seen between two different stars.

We also found that the integrated chromospheric radiative losses
$\phi_\mathrm{{int}}$, normalized to the surface luminosity, show a unique 
trend  for G and K dwarfs when plotted against the $S$ index,
(see Fig. \ref{fi-vs-s}).  This might indicate that 
the same physical processes are heating the stellar chromospheres in 
both cases. We calculated an empirical relationship between $S$
and the energy deposited in the chromosphere, which can be used to
estimate the energetic requirements of a given star knowing its
chromospheric activity level. 

\begin{figure}\centering
\includegraphics[width=7cm]{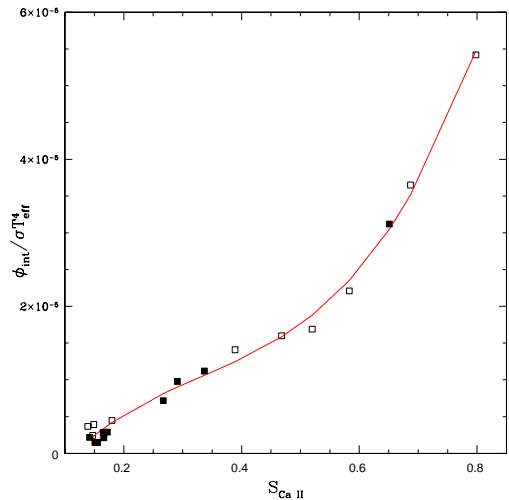}
\caption{Normalized $\phi_\mathrm{{int}}$ versus the $S$ index. The
  squares represent the K star models from \cite{vdm09} and the triangles
  indicate the G star models from \cite{vmc05}.}
\label{fi-vs-s}
\end{figure}



\begin{discussion}

\discuss{STEVE SAAR}  {You mentioned ASAS data in a couple of cases, I would
encourage you to look at the ASAS data for Proxima Centauri.  By an
odd coincidence we have been looking for an x-ray cycle in that
recently and it has what appears to be a good photometric period but
it is more like 8 years rather than the period you quote here.  I have
been wondering if you see a signal in that range. }

\discuss{PABLO MAUAS}  {I was thinking it was time for an update in that star,
since we have four more years of data and we certainly have to use the
ASAS data.  We will have a look at it.}

\discuss{JURGEN SCHMITT}  {Alpha Cen must be perfectly suited for observations
you have -- have you looked at Alpha Cen? }

\discuss{PABLO MAUAS}  {Yes, we have, but we had a problem with the gain of the
pointing screen at the observatory so we had to end those observations
three or four years ago.  But we have a hint of a cycle in the K-star,
Alpha Cen B. }

\discuss{MARK GIAMPAPA} {I think it is great that you are getting cycle
periods for M dwarf stars.  Just a comment on the H-$\alpha$ and Calcium
K.  The models that I have constructed and looked at show that they
are very segregated in their formation regions and so you could have
some, at least in models, disjoint behavior, but I am surprised that
you don't see in observations pretty direct correlation.  But, in any
event I could imagine in the upper atmosphere you could have
flare-like behavior that may not propagate into the lower atmosphere
and that could lead to some deep correlation between Calcium and H~$\alpha$. }

\discuss{PABLO MAUAS}  {Flare activity we usually can detect because we take two
spectra to eliminate cosmic rays, so before we do anything with
our data we check to see if the spectra are more or less the same.
So if you had a flare, that wouldn't happen particularly not for M
stars integration times are very long.  So usually we leave out all
the flares.  For Proxima for example that's hard work because you can
have four or five flares a day.  So usually those extremes are
eliminated from our data.  I think there can be different things for
these different correlations between Calcium and H-$\alpha$.  One can be
filling factor problems.  Perhaps you can explain the same integrated
emission with different filling factors and different contrasts but
they won't give the same relation between H-$\alpha$ and Calcium. The
other one is the orientation - it is not the same if you observe the
star pole-on or from the equator.  So we are exploring that, and preliminary
results are shown in our poster. }

\end{discussion}

\end{document}